\documentstyle[12pt]{article}
\begin{document}
\pagestyle{plain}
\setcounter{page}{1}
\setcounter{footnote}{00}
\baselineskip=18pt 
\def\doublespaced{\baselineskip=\normalbaselineskip\multiply
    \baselineskip by 150\divide\baselineskip by 100}
\begin{flushright}
AMES-HET-98-01\\
NUHEP-TH-98-01\\
\end{flushright}
\vspace{0.2in}

\begin{center}
{\LARGE Top Quark at the Upgraded Tevatron \\ to Probe New Physics }
\footnote{Talk presented by Jin Min Yang at the
Top Quark Working Group at the   
Workshop on Physics at the First Muon Collider 
and at the Front End of a Muon Collider,
Nov. 6-9,  1997, Fermi National Accelerator Laboratory, to appear in 
the proceedings.} 
\vspace{.7cm}

Jin Min Yang $^{a,b,}$\footnote{Address after March: Department of Physics,
Tohoku University, Sendai 980-77, Japan} \\ and 
        \\ A. Datta $^b$, M. Hosch $^b$, C. S. Li $^c$, R. J. Oakes $^a$,\\ 
          K. Whisnant $^b$, Bing-Lin Young $^b$, X. Zhang $^d$ 
\vspace{.5cm}

\it
$^a$ Department of Physics, Northwestern University,
              Evanston, Il 60208\\
         $^b$ Department of Physics and Astronomy, Iowa State University,
              Ames, IA 50011\\
         $^c$ Department of Physics, Peking University, Beijing, China\\
         $^d$ Institute of High Energy Physics, Academia Sinica, Beijing, China
\end{center}
\vspace{2.5cm}

\rm
 \begin{center} Abstract \end{center}

This talk is a brief review of the recent studies on probing new physics 
through single top quark processes and probing exotic top quark decays
 at the upgraded Tevatron. 
\vfill
\eject

\begin{center} \section{Introduction}\end{center}

 The exceedingly heavy top quark is believed 
to be more sensitive to new physics than others.
The upgraded Tevatron will provide a good opportunity to study top quark
properties. Apart from the dominant top pair production which tests the 
top's QCD properties, the single top productions \cite{single}
are also interesting 
to study since they involve the electroweak interaction and can, therefore, 
be used to probe new physics in top's electroweak couplings. Analyses show 
that new physics effects that produce larger than 16\% effect on the single 
top cross section should should be detectable at the upgraded 
Tevatron \cite{heinson}. 
On the other hand, since top events will increase significantly at
the upgraded Tevatron, it is interesting to search for exotic top
decays predicted by new physics models. 

This talk is a brief review of some recent studies on 
 the ability of single top quark production 
at the upgraded Tevatron to probe new physics
as well as the possibility of observing the exotic decay mode 
of the top quark predicted by Minimal Supersymmetric Model (MSSM).
About the SUSY effects in top pair production, we refer
to \cite{pair} and will not discuss them further.  

\begin{center} \section{ Model-independent Analysis for New
      Physics in Single Top Production} \end{center}
 
Since no direct signal of new particles has been observed so far, 
it is very likely that the only observable effects of new physics
at energies not too far
above the SM energy scale could be in the form of new interactions
affecting the couplings of the third-family quarks, and the untested sectors of
the Higgs and
gauge bosons. In this spirit, the new physics effects can be expressed
as non-standard terms in an effective Lagrangian describing the
interactions among third-family quarks, the Higgs and gauge
bosons, which were enumerated in \cite{list1,list2,list3}. In the following
we pick out those which affect single top production at the Tevatron.

Two typical operators 
  which contribute to LEP I and LEP II observables,
$\sigma_{t\bar t}$ and $A_{FB}^t$ at NLC
and $\sigma_{t\bar b}$ at
the Tevatron are \cite{list2}:
\begin{eqnarray}
O_{qW}&=&\left [\bar q_L \gamma^{\mu}\tau^I D^{\nu}q_L
         +\overline{D^{\nu}q_L} \gamma^{\mu}\tau^I q_L\right ]
          W^I_{\mu\nu},\\
O_{\Phi q}^{(3)}&=&i\left [\Phi^{\dagger}\tau^I D_{\mu}\Phi
        -(D_{\mu}\Phi)^{\dagger}\tau^I\Phi\right ]\bar q_L \gamma^{\mu}\tau^I
       q_L.
\end{eqnarray}
Using $1\sigma$ bound of $R_b$ we  obtain the constraints \cite{list2}: 
\begin{eqnarray} 
-0.0080< \frac{4s_Wc_W}{e}\frac{1}{\Lambda^2}\left [
   C_{qW}\frac{c_Wm_Z^2}{4}-
   C^{(3)}_{\Phi q}\frac{vm_Z}{2}\right ]<-0.0023
\end{eqnarray}
Then the effects of $O^{(3)}_{\Phi q}$ are found to be negligibly 
small \cite{list2}, while the effects of $O_{qW}$ are 
\begin{eqnarray}
\begin{array}{lll}
{\rm LEP II}(e^+e^-\rightarrow b\bar b)~&
{\rm NLC}~(e^+e^-\rightarrow t\bar t)~&
{\rm Tevatron}(p\bar p\rightarrow t\bar b+X) \\
 & & \\
2.4{\rm \%}<\frac{\Delta \sigma}{\sigma^0}<8.4 {\rm \%}&
8.6{\rm \%}<\frac{\Delta \sigma}{\sigma^0}<29.8{\rm \%}&
6.9{\rm \%}<\frac{\Delta \sigma}{\sigma^0}<24.0{\rm \%}\\
 & & \\
0.3{\rm \%}<\frac{\delta A_{FB}}{A_{FB}^0}<1.0{\rm \%}&
16.3{\rm \%}<\frac{\delta A_{FB}}{A_{FB}^0}<56.8{\rm \%}& \\
\end{array}
\end{eqnarray}
So the effects of $O_{qW}$ may still be observable at 
LEP II, NLC and the upgraded Tevatron. 

The following dimension-six CP-violating operators 
can give rise to transverse polarization asymmetry of top quark
in single top production ($u+\bar d\rightarrow t+\bar b,~~
\bar u+d\rightarrow \bar t+ b$) at the Tevatron \cite{list3}:
\begin{eqnarray}
\bar O_{qW}&=&i\left [\bar q_L \gamma^{\mu}\tau^I D^{\nu}q_L
         -\overline{D^{\nu}q_L} \gamma^{\mu}\tau^I q_L\right ]
          W^I_{\mu\nu},\\
\bar O_{tW\Phi}&=&i  \left [(\bar q_L \sigma^{\mu\nu}\tau^I t_R) \widetilde\Phi
 -(D^{\mu}\widetilde\Phi)^{\dagger}(\overline {D_{\mu}t_R}q_L)  \right ]
\end{eqnarray}
Introducing the coordinate system
in the top quark ( or top antiquark) rest frame with the unit vectors
$\vec e_z \propto -\vec P_{\bar b}$ and 
$\vec e_y \propto \vec P_u \times \vec P_{\bar b}$.
Transverse polarization asymmetry is defined by \cite{Atwood}
$ A(\hat y)=\frac{1}{2}\left [\Pi(\hat y)-\bar \Pi(\hat y)\right ]$,
where $\Pi(\hat y)$ and $\bar \Pi(\hat y)$ are, respectively,
the polarizations of the top quark and top antiquark in the direction
$\hat y$. The polarizations are given by
\begin{eqnarray}
\Pi(\hat y)=\frac{N_t(+\hat y)-N_t(-\hat y)}{N_t(+\hat y)+N_t(-\hat y)},
~\bar \Pi(\hat y)=\frac{N_{\bar t}(+\hat y)-N_{\bar t}(-\hat y)}
                        {N_{\bar t}(+\hat y)+N_{\bar t}(-\hat y)},
\end{eqnarray}
where $N_t(\pm\hat y)$ $[N_{\bar t}(\pm\hat y)]$ is the number of
$t$($\bar t$) quarks polarized in the direction $\pm\hat y$.

Assuming $m_t=175$ GeV, we obtain the asymmetry at hadron level as \cite{list3}
\begin{eqnarray}
A(\hat y)&=&\left \{
\begin{array}{ll}
-0.41\frac{C_{qW}-2C_{tW\Phi}-g_2C_{Dt}/2}{(\Lambda/1 {\rm ~TeV})^2}
                                           &~~{\rm at}~\sqrt s=2~{\rm TeV}\\
-0.84\frac{C_{qW}-2C_{tW\Phi}-g_2C_{Dt}/2}{(\Lambda/1 {\rm ~TeV})^2}
                                           &~~{\rm at}~\sqrt s=4~{\rm TeV}
\end{array}\right.
\end{eqnarray}
Assume an observable level of ten percent of this asymmetry,
the upgraded Tevatron will probe
\begin{eqnarray}
\frac{ C_{qW}-2C_{tW\Phi}-g_2C_{Dt}/2 }{(\Lambda/1 {\rm ~TeV})^2}~~
{\rm to} ~~\left \{
\begin{array}{ll}  1/4 & ~{\rm for}~~\sqrt s=2~ {\rm TeV}\\
		   1/8 & ~{\rm for}~~\sqrt s=4~ {\rm TeV}\\
\end{array} \right.
\end{eqnarray}
This means that with a  new physics scale at the order of
1 TeV, the further upgraded Tevatron can probe the coupling strength
down to the level of 0.1.

\begin{center}\section{ Probing SUSY in Single Top Production }
\end{center}

In the R-parity Conserving MSSM,
we found that within the  allowed range of squark and gluino masses
the supersymmetric QCD corrections can enhance the cross section
by a few percent \cite{Li1}. 
The Yukawa corrections \cite{Li1} to single top quark production 
at the Tevatron can amount to more 
than a 15\% reduction in the production cross section relative to the tree 
level result in the general two-Higgs-doublet model,
and a 10\% enhancement in the minimal supersymmetric model
for the smallest allowed $\tan\beta~ (\simeq 0.25)$. 
The supersymmetric electroweak corrections \cite{Li1} to the cross section
are at most a few percent for $\tan \beta>1$, but can exceed 
10\% for $\tan\beta<1$.
So the combined effects of SUSY QCD, SUSY EW, and the Yukawa couplings
in the R-parity Conserving MSSM
can exceed 10\%  for the smallest allowed $\tan\beta~ (\simeq 0.25)$ 
but are only a few percent for $\tan\beta >1$. 

In the R-parity violating MSSM,  
the processes induced by R-violating couplings are \cite{datta,bob}
\begin{eqnarray}
\lambda^{\prime}:& ~~~& 
u\bar d\rightarrow \tilde l \rightarrow t+\bar b, {\rm ~(s-channel)}\\
\lambda^{\prime\prime}: & ~~~& 
u\bar d \rightarrow t+\bar b, {\rm ~(t-channel)}\\
\lambda^{\prime\prime}: & ~~~& c d \rightarrow \tilde s \rightarrow t b,
        {\rm ~(s-channel)}\\
\lambda^{\prime\prime}: & ~~~& c s \rightarrow \tilde d \rightarrow t b,
        {\rm ~(s-channel)}
\end{eqnarray}
Their signature are an energetic charged lepton, missing $E_{T}$, 
and double $b$-quark jets.
The backgrounds are (1) $q\bar q' \rightarrow W^* \rightarrow t\bar b$,
(2) the quark-gluon process
                $qg\rightarrow q't\bar b$ with a W-boson
                as an intermediate state in either the t-channel
                or the s-channel of a subdiagram;
(3) processes involving a b-quark in the initial state,
                $bq (\bar q) \rightarrow tq'(\bar q')$ and $gb\rightarrow tW$;
(4) $Wb\bar b$;
(5) $Wjj$; and 
(6) $t\bar t\rightarrow W^-W^+b\bar b$.
For the upgraded Tevatron (LHC), the basic cuts are
$p_T^l\ge 20 \rm{~GeV}$, $p_T^b\ge 20(35) \rm{~GeV}$,
$p_T^{\rm miss}\ge 20(30) \rm{~GeV}$, $\eta_{b},~\eta_{l} \le 2.5(3)$ and
$\Delta R_{jj},~\Delta R_{jl} \ge 0.4$.
Also we required reconstructed top quark mass $M(bW)$ to lie
within the mass range $\left \vert M(bW)-m_t\right \vert <30~{\rm GeV}$,
which can reduce the backgrounds $Wb\bar b$ and $Wjj$ efficiently.
The number of signal events required for discovery of a signal is
$S \ge 5 \sqrt{B}$. 

 For the s-channel process 
$u+\bar d\rightarrow \tilde l \rightarrow t+\bar b$ induced 
by $\lambda^{\prime}$, the histogram of the differential cross section
versus the invariant mass of the $t\bar b$ system over the bin size of
10 GeV is shown in Fig.2 of Ref. \cite{datta}.
The resonance behavior is already manifested.
Because of their narrow widths, for each slepton the contributions of
the $\lambda^{\prime}$-couplings are negligible for a couple of bins away
from the resonance.  This will help to
identify the signal of the slepton production. 
The value of $\lambda'_{111}\lambda'_{133}+\lambda'_{211}\lambda'_{233}
+\lambda'_{311}\lambda'_{333}$ versus the slepton mass for
$u \bar d \rightarrow \tilde l \rightarrow t \bar b$
to be observable under the criteria $ S \ge 5 \sqrt{B}$
is shown in Fig.4 of Ref. \cite{bob}, which show that
the LHC can do better than the upgraded Tevatron
in further probing the couplings, especially for higher mass sleptons.  

For the s-channel process $ c d \rightarrow \tilde s \rightarrow t b$
induced by $\lambda^{\prime\prime}$ couplings, the value of 
$\lambda''_{212} \lambda''_{332}$ versus strange-squark mass 
for it to be observable under the criteria $ S \ge 5 \sqrt{B}$ is
shown in Fig.1 of Ref.\cite{bob}, which 
show that both the LHC and the upgraded Tevatron can efficiently probe
the relevant couplings, and the LHC serves a more powerful probe than
the upgraded Tevatron. 

 For the s-channel process $c s \rightarrow \tilde d \rightarrow t b$
induced by $\lambda^{\prime\prime}$, 
The value of $\lambda''_{212} \lambda''_{331}$
versus down-squark mass for it
to be observable under the criteria $ S \ge 5 \sqrt{B}$ is shown in Fig.3
of Ref. \cite{bob}, which show that this  process 
cannot be probed as efficiently  as  $c d \rightarrow \tilde s \rightarrow t b$
because of the relative suppression of the strange quark structure function
compared to the valence down quark.

\begin{center}\section{Searching for Exotic Top Decay Modes }
\end{center}

The FCNC Decays in the SM, 
 R-conserving MSSM \cite{Li2} and R-violating MSSM \cite{Yang}
are given by
\begin{eqnarray} 
\begin{array}{lccc} 
~                 & {\rm SM} & {\rm~~~ MSSM~~~} &\not \! R {\rm ~ MSSM}\\
B(t\rightarrow cg)~~~~     & 10^{-10} & 10^{-6}            & 10^{-3} \\  
B(t\rightarrow c\gamma)& 10^{-12} & 10^{-8}            & 10^{-5} \\
B(t\rightarrow cZ)     & 10^{-12} & 10^{-8}            & 10^{-4} \\
B(t\rightarrow ch)     & 10^{-7}  & 10^{-5}            & ~       
\end{array}
\end{eqnarray}
The FCNC top decays in R-violating MSSM might be observable
at the upgraded Tevatron since for integrated luminosity of 10 (100) fb$^{-1}$,
the detection sensitivity is \cite{Han}
$Br(t \rightarrow cg)      \simeq 5\times 10^{-3}(1\times 10^{-3})$,
$Br(t \rightarrow c\gamma) \simeq  4\times 10^{-4}(8\times 10^{-5})$ and
$Br(t \rightarrow cZ)      \simeq  4\times 10^{-3} (6\times 10^{-4})$.

Let us look at the top decay to light stop,
 $t\rightarrow \tilde t_1 \tilde\chi^0_1$, 
in the framework of R-conserving MSSM with the lightest neutralino 
being the LSP.  the parameters involved in
$\Gamma(t\rightarrow \tilde t_1 \tilde \chi^0_1)$ are
$M_{\tilde t_1},  M_2, M_1, \mu, \tan\beta$.
In the region of parameter space allowed by $R_b$ data and the
$ee\gamma\gamma+{\large \not} \! E_T$ event\cite{Kane},
we obtain \cite{Matt} 
$0.07\le B(t\rightarrow \tilde t_1\tilde \chi^0_1)\le 0.50$. 
The dorminant decay of a light stop is 
$\tilde t_1 \rightarrow c\tilde \chi^0_1 $.
This will give a new final state in $t\bar t$ production: 
$t\bar t \rightarrow Wb\bar c\tilde \chi^0_1 \tilde \chi^0_1$. 
Its  signature is  an energetic charged lepton,
one $b$-quark jet, one light $c$-quark jet, plus missing $E_{T}$ from
the neutrino and the unobservable ${\chi^0_1}^\prime$s.
The potential SM backgrounds are (1) $bq (\bar q) \rightarrow tq'(\bar q')$,
(2) $q\bar q' \rightarrow W^* \rightarrow t\bar b$;
(3) $Wb\bar b$; (4) $Wjj$; (5) $t\bar t\rightarrow W^-W^+b\bar b$;
(6) $gb\rightarrow tW$ and (7) $qg\rightarrow q't\bar b$.
Besides the basic cuts we impose a cut on  transverse mass
$m_T = \sqrt{ (P_T^l+P_T^{\rm miss})^2
- (\vec P_T^l+\vec P_T^{\rm miss})^2}>90$ GeV.
Then we found (1) this final state is unobservable at 
Run 1 with $\sqrt s=1.8$ TeV and $L=0.1$ fb$^{-1}$,
(2) Run 2 with $\sqrt s=2$ TeV and $L=10$ fb$^{-1}$ can 
either discover this final state or provide the additional strong 
constraint given approximately by 
$M_{\tilde t_1}-M_{\tilde \chi^0_1} < 6 ~{\rm GeV}$.

If charged
Higgs is light enough, $t \rightarrow H^+ b$
is also possible; we refer to \cite{sola} for its 
phenomenological implications at Tevatron.

\begin{center} \section{Conclusion }
\end{center}

(1) Single top quark processes at upgraded Tevatron can be meaningfuly 
used to probe new physics; (2) The FCNC top decays 
$t\rightarrow c V$ and 
top decay to light stop $t\rightarrow \tilde t_1 \tilde\chi^0_1$ 
predicted by R-violating MSSM might be observable at upgraded Tevatron,
else further constraints can be set on the relevant couplings.


\vspace{1cm}

\begin{thebibliography}{99}
\frenchspacing

\bibitem{single}  S. Willenbrock and D. Dicus, {\it Phys. Rev.} {\bf D34}, 155 (1986);
                S. Dawson and S. Willenbrock, {\it Nucl. Phys.} {\bf B284}, 449 (1987);
                C.-P. Yuan, {\it Phys. Rev.} {\bf D41}, 42 (1990);
                F. Anselmo, B. van Eijk and G. Bordes, 
                {\it Phys. Rev.} {\bf D45}, 2312 (1992);
                R. K. Ellis and S. Parke, {\it Phys. Rev.} {\bf D46},3785 (1992);
                D. Carlson and C.-P. Yuan, {\it Phys. Lett.} {\bf B306},386 (1993);
                G. Bordes and B. van Eijk, {\it Nucl. Phys.} {\bf B435}, 23 (1995);
                A. Heinson, A. Belyaev and E. Boos, hep-ph/9509274. 
                S. Cortese and R. Petronzio, {\it Phys. Lett.} {\bf B306}, 386 (1993). 
                T. Stelzer and S. Willenbrock, {\it Phys. Lett.} {\bf B357}, 125 (1995). 
                M. Smith and S. Willenbrock,  {\it Phys. Rev.} {\bf D54}, 6696 (1996);
                S. Mrenna and C.-P. Yuan, hep-ph/9703224; 
                T. Tait and C.-P. Yuan, hep-ph/9710372.
\bibitem{heinson} A. P. Heinson, hep-ex/9605010. 
\bibitem{pair}  C. S. Li, B. Q. Hu, J. M. Yang and C. G. Hu, {\it Phys. Rev.} {\bf D52},
                 5014 (1995);
    J. M. Yang and C. S. Li, {\it Phys. Rev.} {\bf D52}, 1541 (1995);
    J. Kim, J. L. Lopez, D. V. Nanopoulos and R. Rangarajan, 
      {\it Phys. Rev.} {\bf D54}, 4364 (1996);
    J. M. Yang and C. S. Li, {\it Phys. Rev.} {\bf D54}, 4380 (1996);
    C. S. Li, H. Y. Zhou, Y. L. Zhu, J. M. Yang, {\it Phys. Lett.} {\bf B379}, 135 (1996);
    S. Alam, K. Hagiwara and S. Matsumoto, {\it Phys. Rev.} {\bf D55}, 1307 (1997);
    Z. Sullivan, {\it Phys. Rev.} {\bf D56}, 451 (1997);
    W. Hollik, W. M. Mosle and D. Wackeroth, hep-ph/9706218.
\bibitem{list1}  G. J. Gounaris, D. T. Papadamou and F. M. Renard, hep-ph/9609437.
\bibitem{list2}  K. Whisnant, J. M. Yang, B.-L. Young and X. Zhang, 
                {\it Phys. Rev.} {\bf D56}, 467 (1997). 
\bibitem{list3}  J. M. Yang and B.-L. Young,  {\it Phys. Rev.} {\bf D56}, 5907 (1997). 
\bibitem{Atwood} D. Atwood, S. B. Shalom, G. Eilam and A. Soni,{\it Phys. Rev.}
                 {\bf D54}, 5412  (1996). 
\bibitem{Li1} C. S. Li, R. J. Oakes and J. M. Yang, {\it Phys. Rev.} {\bf D55}, 1672 (1997);
              {\it Phys. Rev.} {\bf D55}, 5780 (1997); hep-ph/9706412.
\bibitem{datta} A. Datta, J. M. Yang, B.-L. Young and X. Zhang,
                {\it Phys. Rev.} {\bf D56}, 3107 (1997);
\bibitem{bob}  R. J. Oakes, K. Whisnant, J. M. Yang, B.-L. Young and X. Zhang,  
              hep-ph/9707477.
\bibitem{Li2} C. S. Li, R. J. Oakes and J. M. Yang, {\it Phys. Rev.} {\bf D49}, 293 (1994);
              J. M. Yang and C. S. Li, {\it Phys. Rev.} {\bf D49}, 3412 (1994);
              G. Couture, C. Hamzaoui and H. Konig, {\it Phys. Rev.} {\bf D52}, 171(1995);
                J. L. Lopez, D. V. Nanopoulos and R. Rangarajan, hep-ph/9702350;
                G. Couture, M. Frank and H. Konig, hep-ph/9704305;
                G. M. de Divitiis, R. Petronzio and L. Silvestrini,
                                                hep-ph/9704244.
\bibitem{Yang} J. M. Yang, B.-L. Young and X. Zhang, hep-ph/9705341.
\bibitem{Han} T. Han, R. D. Peccei, and X. Zhang, {\it Nucl. Phys.} {\bf B454},
                527 (1995);
                T. Han, K. Whisnant, B.-L. Young, and X. Zhang,
                {\it Phys. Rev.} {\bf D55}, 7241 (1997);
                {\it Phys. Lett.} {\bf B385}, 311 (1996).
\bibitem{Kane} S. Ambrosanio, G. L. Kane, G. D. Kribs, S. P. Martin,
                  and S. Mrenna, {\it Phys. Rev. Lett.} {\bf 76}, 3498 (1996);
                  S. Dimopolous, M. Dine, S. Raby and S. Thomas,
                  {\it Phys. Rev. Lett.} {\bf 76}, 3502 (1996);
                   D. Garcia and J. Sola, {\it Phys. Lett.} {\bf B357}, 349 (1995).
\bibitem{Matt} M. Hosch, R. J. Oakes, K. Whisnant,
               J. M. Yang, B.-L. Young and X. Zhang, hep-ph/9711234;
               S. Mrenna and C. P. Yuan, {\it Phys. Lett.} {\bf B367}, 188 (1996).
\bibitem{sola} J. Guasch and J. Sola, hep-ph/9707535.
\end{thebibliography}
\end{document}